\documentclass[]{aa} 

\usepackage{graphicx} 
\usepackage{amssymb} 
\usepackage{epsfig} 
\usepackage{natbib} 

\def\bb{{\sc bb}}

\def\nh{{$N_{\rm H}$}} 
 
\def\bb{{\sc bb}} 
\def\I{{\em INTEGRAL}} 
\def\SLX{SLX~1737-282} 
\def\be{\begin{equation}} 
\def\ee{\end{equation}}

\begin{document} 
 
\title{Intermediate long X-ray bursts from the ultra-compact
  binary  candidate \SLX} 
 
\author{M. Falanga,\inst{1,2} 
J. Chenevez,\inst{3}  
A. Cumming,\inst{4}
E. Kuulkers,\inst{5}
G. Trap,\inst{1,6}  
A. Goldwurm,\inst{1,6}
} 
 
\offprints{M. Falanga} 
\titlerunning{Intermediate long bursts from SLX~1737-282} 
\authorrunning{M. Falanga, et al.} 
 
\institute{CEA Saclay, DSM/DAPNIA/Service d'Astrophysique 
     (CNRS FRE 2591), 91191, Gif sur Yvette, France
     \email{mfalanga@cea.fr} 
\and AIM - Unit\'e Mixte de Recherche CEA - CNRS -
  Universit\'e Paris 7, Paris, France
\and National Space Institute, Technical University of Denmark, 
           Juliane Maries Vej 30, 2100 
           CopenhagenØ, Denmark 
\and Physics Department, McGill University, 3600 rue University,
     Montreal QC, H3A 2T8, Canada
\and ISOC, ESA/ESAC, Urb. Villafranca del Castillo, 
     P.O. Box 50727, E-28080 Madrid, Spain 
\and Universit\'e Paris Diderot-Paris 7 et Observatoire de Paris,
Laboratoire APC, Paris, France
}

\abstract 
{} 
{The low persistent flux X-ray burster source \SLX\ is classified as an
  ultra-compact binary candidate. We compare the data on \SLX\ with the
  other similar objects and attempt to derive constraints on the
  physical processes responsible for the formation of intermediate long bursts. 
}  
{Up to now only four bursts, all with duration
  between  $\simeq15-30$ minutes, have been
recorded for \SLX. The properties of three of these intermediate long X-ray
bursts observed by \I\ are investigated and compared to other burster
sources.  The broadband spectrum of the persistent emission in the
3--100 keV energy band is studied with the \I\ data.  
}   
{The persistent emission is measured to be $0.5\%$ Eddington luminosity.
From the photospheric radius expansion observed during at least one burst
  we derive the source distance at 7.3 kpc assuming a pure helium
  atmosphere. The observed intermediate long burst 
properties from \SLX\ are consistent with helium ignition at the
column depth of 5-8 $\times 10^{9}$ g cm$^{-2}$ and a burst energy release of
$\sim10^{41}$ erg. The apparent recurrence time of $\simeq 86$ days 
between the intermediate long bursts from \SLX\ suggests
a regime of unstable burning of a thick, pure helium layer slowly accreted from
a helium donor star.  
}
{}

 
\maketitle 
 
\section{Introduction} 
\label{sec:intro} 

Since the first complete Galactic plane scan by Uhuru in the 70's,
X-ray missions have revealed that most
neutron star low-mass X-ray binary (LMXB) systems  exhibit type I X-ray
bursts \citep{liu07}. Type I bursts are 
thermonuclear explosions on the surface of accreting neutron stars (NS)
triggered by unstable hydrogen or helium burning.  
They are typically characterized by a fast rise time of $\sim 1$ s,
exponential-like decays with durations ranging from seconds to
minutes, and recurrence times from a few hours to days \citep[see,
  e.g.,][for reviews]{lewin93,sb06}. 
Several thousand bursts have been observed
to date \citep[see, e.g.,][]{cornelisse03,galloway06,chelovekov06}. 

Only on a few occasions, for 10\% of the bursters, type I
X-ray bursts have shown decay times ranging between ten and a few tens of
minutes 
\citep[e.g.,][]{Swank,Hoff,kul02,intZ02,molkov05,intZ05,chenevez06,chenevez07}.
Such intermediate long bursts have durations and energy releases
($\sim10^{41}$ erg),  
intermediate between usual type I X-ray bursts, and so-called superbursts 
lasting more than an hour \citep[e.g.,][]{ek04}.  
Fifteen superbursts have been detected from ten sources to date
\citep[][and references therein]{ek04,intZ04,remillard05,ek05,keek07}. 
It is thought that unstable carbon burning \citep{wt76,sb02} in an
ocean of heavy nuclei \citep{cb01}, possibly combined with
photo-disintegration-triggered nuclear energy release \citep{s03}, is
responsible for most superbursts.  

The mechanisms driving the intermediate long bursters at very low
persistent luminosity have been the subject of recent investigations 
\citep{Peng, CoopN}, suggesting that thermally-unstable hydrogen ignition 
results in sporadic energetic helium bursts
in a mixed hydrogen and helium environment. However, intermediate long helium
bursts have also been proposed to be observed 
at low pure helium accretion rates \citep[e.g.,][]{intZ05,intZ07,cum06}. 

The estimated fraction of intermediate long bursts or superbursts
among the whole type I burst population is only 0.3--0.4\%. This
shows that intermediate long bursts are very scarce events. In this
paper, we report the identification of three additional intermediate long
bursts from \SLX. As far as we
know, this makes \SLX\ the only burster source, which exclusively exhibits
intermediate long bursts (four bursts all longer than 15 minutes
observed so far, see \citet{intZ02} for the first observed burst for this
  source, and \citet{sguera07a} for the last reported burst).  
The present analysis concentrates on the properties of the three intermediate
long bursts observed with \I. We compare the observed intermediate
long bursts with different burst types.

\subsection{The source \SLX} 
\label{sec:source} 

\citet{skinner87} discovered \SLX\ in 1985 with the {\it Spacelab-2}
observatory as a low persistent X-ray emission source in the
energy range $3-30$ keV. Since then, this source has been observed serendipitously during
 monitoring programs of the  Galactic center by different X-ray
observatories at the flux level of $(5-10) \times 10^{-11}$ erg
cm$^{-2}$ s$^{-1}$ 
\citep{skinner87,sakano02,intZ02,tomsick07}. The most accurate
position of the source
has been provided by {\it Chandra} at $\alpha_{\rm J2000} = 17^{\rm
  h}40^{\rm m}42\fs83$ and $\delta_{\rm J2000} = -28{\degr}18\arcmin08\farcs4$
 with an estimated accuracy of $0\farcs6$ \citep{tomsick07}. 

From its spectral analysis and flux variability the
system \SLX\ was more likely classified as a low-mass X-ray binary
\citep{sakano02}.  
On March 17, 2000, the discovery of a $\approx15$ min intermediate long type I
X-ray burst,  with the
Wide Field Camera onboard the {\it BeppoSAX} observatory,
led to the classification of \SLX\ as a NS low-mass X-ray
binary system \citep{intZ02}. The source distance was estimated to be
between 5 and 8 kpc from the  burst peak flux during the radius
expansion of about $6\times10^{-8}$ erg cm$^{-2}$ s$^{-1}$ \citep{intZ02}.

The burst detected with {\it BeppoSAX} remained the only one known
 from this system until the launch of the \I\ satellite. A large exposure
time spent by \I\ on observations of the Galactic center region allowed
us to detect three additional intermediate long bursts from \SLX.  All
three bursts had a duration of around 20--30 minutes. Here we study these
additional intermediate long bursts.

\section{Data Analysis and results} 
\label{sec:integral} 
 
Three intermediate long bursts have been observed by \I/JEM-X and
IBIS/ISGRI cameras \citep[][]{w03,lund03,u03,lebr03}, on March 9, 2004 (burst
1), April 11, 2005 (burst 2) 
and April 2, 2007 (burst 3), respectively.
Bursts 2 and 3 were also detected with the \I\ Burst
Alert System (IBAS) software \citep{mereghetti03},
which is dedicated to the real-time discovery and localization of
gamma-ray bursts, transient X-ray sources, and bursts in the IBIS/ISGRI
data stream. 
One of IBAS operation modes, running on the
15--40 keV energy interval and 10 s timescale, is particularly
well suited to detecting type I X-ray bursts. Typically, the bursts are
localized with a $\sim3'$ uncertainty.  

We performed the data reduction with the standard 
Offline Science Analysis (OSA) software version 7.0.
When performing analysis on a list of events collected by 
a coded mask telescope it is convenient, to increase the 
signal to noise ratio for a given source,
to select events that were recorded by the parts of the detector 
illuminated by that source through the transparent elements
of the mask. For weak sources it has been found that the best 
results are obtained when selection includes also partially-illuminated 
detector pixels with an illumination fraction higher than about 0.4-0.5.
Light curves and spectra must also be corrected 
for off-axis efficiency reduction due to partial modulation and variation 
of mask hole opacity with incidence angle, using the appropriate correction
factor computed for the given source \citep{goldwurm03}.

The off-axis corrected burst light curves are based on events 
selected according to the detector illumination pattern for \SLX; for 
ISGRI we used an illumination threshold 
of 0.4 for the energy range 20--60 keV, for JEM-X we used the source
events in the 3--20 keV energy range. 
For the persistent spectral analysis, we extracted the \I\ data for all
pointings within $5^{\circ}$ (JEM-X) and $9^{\circ}$ (ISGRI) of the
source position  for a total effective exposure of about 2.3 Ms and 4 Ms,
respectively (March 2003 to October 2006 and from the April 2007
Galactic center observation).

Since the source is not detected in single pointings it was
necessary to increase the sensitivity by combining the observations
to accumulate as much exposure time as possible. Therefore,  
to study the weak persistent X-ray emission (see 
Sec. \ref{sec:persitent}), we derived the JEM-X and ISGRI spectra from
total mosaic images in four energy bands for JEM-X (3--20 keV)
and five energy bands for ISGRI (20--100 keV).    
We applied a systematic error of 3\% to combined JEM-X and ISGRI spectrum, 
which corresponds to the current uncertainties in the response matrices.
All uncertainties in the spectral parameters are given at a 90\% 
confidence level for single parameters.

\subsection{Persistent emission}  
 \label{sec:persitent} 

\begin{figure}
\centerline{\epsfig{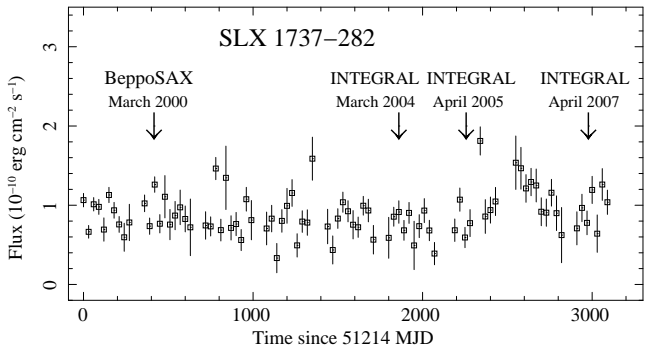}}
\caption{{\it RXTE} bulge observations light curve for
  \SLX\ averaged over 30-day intervals from February 5, 1999 to
  August 12, 2007.
The {\it RXTE}/PCA count rate has been converted into flux using the 2--10
  keV band Crab flux of $1.2\times10^{-10}$ erg cm$^{-2}$ s$^{-1}$
  for 60 cts s$^{-1}$ per 5 PCUs \citep{intZ02}.
The times of the first burst, {\it BeppoSAX} observation, and the last
  three bursts, this work, are indicated.}
\label{fig:lc-rxte}
\end{figure}

In Fig. \ref{fig:lc-rxte} we plot the 2--10 keV persistent emission for 
\SLX\ obtained with {\em RXTE} bulge
observations\footnote{http://lheawww.gsfc.nasa.gov/users/craigm/\\
galscan/main.html} \citep{swankmark01}, which shows that the source is
a weak persistent X-ray source.   
Since the persistent emission of \SLX\ is more or less stable over the whole
available data set, its average broadband spectrum has been obtained by
combining the JEM-X (3--20 keV) and IBIS/ISGRI (20--100 keV)
observations. Note that neither JEM-X nor ISGRI detected the
persistent emission 10 ks before and after the burst intervals. 
The energy range covered by JEM-X does not allow us to constrain  the
interstellar hydrogen column density, {\nh}.  
We fixed in all our spectral fits \nh\ to $1.9\times10^{22}$
cm$^{-2}$, the value found from {\em BeppoSAX} and {\it ASCA} \citep{intZ02}. 

\begin{figure}[htb] 
\centerline{\epsfig{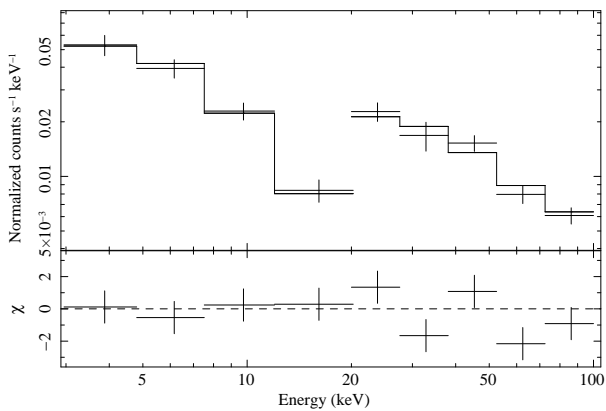}}
\caption 
{{\it INTEGRAL}/JEM-X (3--20 keV) and IBIS/ISGRI (20--100 keV)
  averaged spectrum of \SLX\ persistent emission. The best fit is   
 obtained with a simple absorbed power-law model.
 The upper panel shows the data and the best fit 
 model, whereas in the lower panel we plot the residuals from this 
 fit.}  
\label{fig:spec} 
\end{figure}

The joined JEM-X/ISGRI 3--100 keV broadband spectrum 
is best fitted with a photoelectrically-absorbed power-law model, 
with $\chi^{2}{\rm /d.o.f.} = 8.5/7$, and a photon index
$\Gamma=2.1\pm0.1$.  
The 3--100 keV unabsorbed flux is $(1.3\pm0.15)\times10^{-10}$ erg cm$^{-2}$
s$^{-1}$, which extrapolated in the 0.1--100 keV band is found to be
$(3.0\pm0.45)\times10^{-10}$ erg cm$^{-2}$ s$^{-1}$. 
A multiplicative factor was included in the fit to 
take account of the uncertainty in the cross-calibration of the instruments. 
The factor was fixed at 1 for the JEM-X data and the 
normalizations of the ISGRI data were found within $1.2\pm0.1$.
The best fit on the count rate spectrum and the residuals from
  this fit are shown in Fig. \ref{fig:spec}.

\subsection{Burst light curves} 
\label{sec:lcburst} 
 
\begin{figure*}[htb]
\centerline{\epsfig{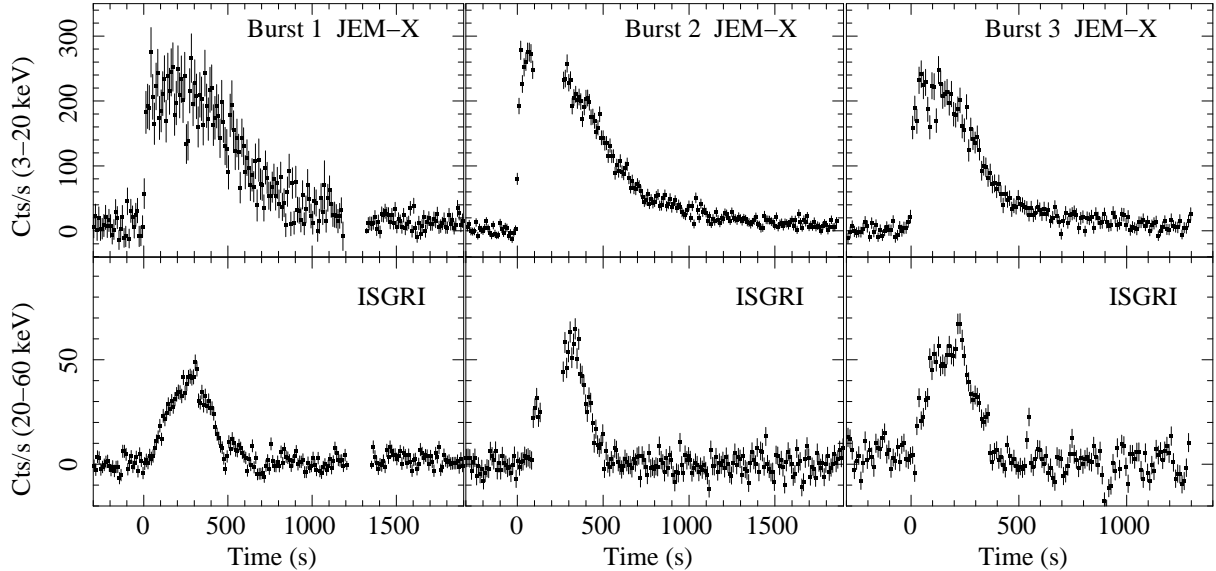}}
\caption 
{The intermediate long type I X-ray bursts detected from \SLX\ on March
9, 2004 (burst 1); April 11, 2005 (burst 2); and April 2, 2007 (burst
3), respectively.  
The time T$_{0}$'s expressed in UTC corresponds to $17^{\rm h}18^{\rm
  m}49^{\rm s}$, $08^{\rm h}10^{\rm
  m}28^{\rm s}$, and $05^{\rm h}57^{\rm m}8^{\rm s}$, respectively.  
The JEM-X (3--20 keV) and ISGRI (20--60 keV) net light curve
(background subtracted) are shown with a time bin of 10 s.  
The data gap, in bursts 1 and 2, is due to a two-minute slew interval between
two \I\ stable  pointings.}  
\label{fig:burst1} 
\end{figure*}

\begin{figure}[htb]
\centerline{\epsfig{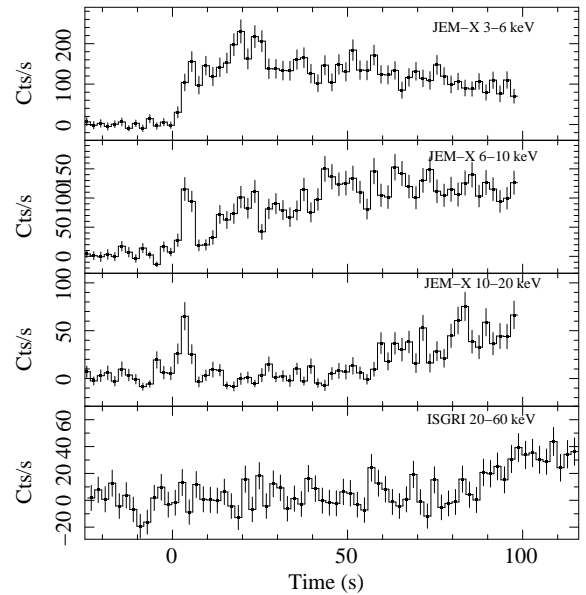}} 
\caption 
{Temporal profile of burst 2 (see Fig. \ref{fig:burst1}) measured
 with JEM-X and ISGRI at different energy bands. At T$_{0}$ a
 soft precursor can be observed at different energy bands. The net
  light curve are shown with a time bin of 2 s.
}  
\label{fig:burstzoom} 
\end{figure}

In Fig. \ref{fig:burst1}, 
we show the JEM-X 3--20 keV (upper panels) and ISGRI 20-60 keV (lower
panels) light curves for bursts 1, 2, and 3, respectively.  
Bursts 1 and 2 (see Fig. \ref{fig:burst1}) were  each observed over two
consecutive stable pointings with a 2 minute slew in
between, during which no data are available.
Note that burst 1 was $4.9^{\circ}$ off-axis during the first
pointing, therefore for JEM-X the errors on the light curve are a
factor $\sim 2$ higher compared to burst 2 and 3. 
Figure \ref{fig:burstzoom} shows that burst 2 started with a short
burst-like  soft event (``precursor'') that lasted $\sim 6$ s,
and which was more powerful in the soft energy bands.
After the precursor, at higher energies, the source returned
to the persistent flux level. Such behavior is typical for bursts with
an extreme photospheric radius expansion and can be interpreted in
terms of cooling of the NS photosphere during its expansion. After
this event, the source's flux rose again to maximum, depending on the 
energy bands: a relatively quick increase of the intensity was first
observed at low energies and then became gradually
visible at higher energies. This can also be seen between JEM-X and
ISGRI light curves on Fig. \ref{fig:burst1}. 
Such behavior reflects a strong change in the hardness and is also
typical for bursts with photospheric radius expansion \citep[see,
  e.g.,][]{lewin93}.  

The start time for each burst was determined when the intensity rose to 10\% 
of the peak above the persistent intensity level. 
The rise time is defined as the time between the start of the burst and 
the time at which the intensity reached 90\% of the peak burst intensity.
For all three bursts it was $2\pm1$~s; the e-folding decay time, 
determined over the time after the plateau, was $\tau =298\pm22$~s,
$\tau =321\pm10$~s, and $\tau=270\pm14$~s, respectively for bursts 1,
2, and 3. The total duration, i.e., from the burst start time back to
the persistent flux level in the 3--20 keV band was 25, 30, and 20
minutes, respectively.  

\subsection{X-ray burst spectra} 
\label{sec:burstspec} 
  
\begin{figure}[htb]
\centerline{\epsfig{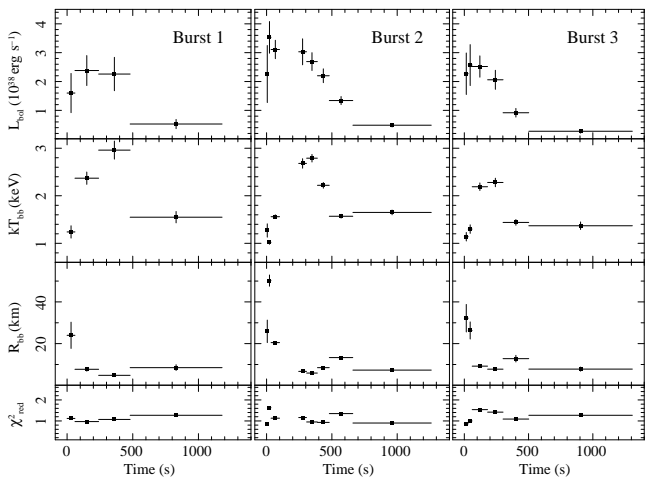}} 
\caption 
{The spectral evolution measured by JEM-X and ISGRI during bursts 2
  and 3, respectively; bolometric luminosity at 7.3 kpc (assuming a
  helium burst), radius of the  
  photosphere and its temperature during the bursts are obtained using the \bb\
  model for the spectral fit. $\chi^{2}_{\rm red}$ values are shown in
  the bottom.   
}  
\label{fig:fig5} 
\end{figure}

For the spectral analysis of the bursts we used JEM-X/ISGRI data in the
3--20 keV and 20--60 keV bands, respectively. 
We performed time-resolved spectral analysis. The net burst
spectra are well fitted by a simple \bb\ model. 
The bolometric luminosity, the inferred \bb\
temperature, $kT_{\rm bb}$, and apparent \bb\ radius, $R_{\rm bb}$, 
at 7.3~kpc (see Sec. \ref{sec:burst_pers_distance}) are reported in
Fig. \ref{fig:fig5}. In Table \ref{table:spec1} we report the burst
parameters.  

The bursts fluence are obtained from the bolometric fluxes, $F_{\rm bol}$, 
extrapolated in the 0.1--100 keV energy range over the respective burst durations.
The peak fluxes, $F_{\rm peak}$, are derived from the 3--20 keV light
curve peak count rates with 2 s time resolution and renormalized for
the bolometric energy range. 

\section{Discussion}  
\label{sec:discussion} 

\subsection{Bursts, persistent flux, and source distance}
\label{sec:burst_pers_distance}
 
The present bursts are well described by a simple \bb\ model 
representing the thermal emission from the NS surface, 
which is a common observed property of type I X-ray bursts 
\citep[see, e.g.,][]{galloway06}.

\begin{table}[htb] 
\caption{Burst parameters.} 
\label{table:spec1}
\begin{center} 
\renewcommand{\footnoterule}{}
\begin{tabular}{llll} 
\hline \hline
\noalign{\smallskip} 
 & Burst 1 & Burst 2 &   Burst 3     \\
\hline 
\noalign{\smallskip}
$F_{\rm peak}^{a}$ ($10^{-8}$ erg cm$^{-2}$ s$^{-1}$) &$ 4.0\pm0.8$ &  $6.0\pm0.5$ & $5.7\pm0.5$ \\
$f_{\rm b}^{b}$ ($10^{-5}$erg cm$^{-2}$)  & $1.1\pm0.03$ & $1.94\pm0.02$ & $1.6 \pm0.02$ \\
$\tau \equiv f_{\rm b}/F_{\rm peak}$ (sec)  & $275\pm55$ &  $323\pm27$  & $281\pm25$ \\
$\gamma  \equiv F^{c}_{\rm pers}/F_{\rm peak}$ $(10^{-3})$ & $7.5\pm2$ &  $5.0\pm0.8$  &  $5.3\pm0.9$ \\
\hline
\end{tabular}
\end{center} 
\small $^{a}$ Unabsorbed flux (0.1--100 keV).
\small $^{b}$ Fluence. 
\small $^{c}$ Using the unabsorbed persistent flux $F_{\rm
  pers}=(3\pm0.45)\times 10^{-10}$ erg cm$^{-2}$ s$^{-1}$ (0.1--100 keV). 
\end{table} 

The light curve and spectral analysis of burst 2 show evidence for
photospheric radius expansion 
indicating that its bolometric peak luminosity reached the Eddington
limit. Assuming the Eddington luminosity for a helium burst, $L_{\rm
 Edd}\approx\,3.8\times 10^{38}$ ergs$^{-1}$, as empirically derived
by \citet{kul03}, we calculate the source distance to be 7.3 kpc. 
For comparison, the theoretical
value \citep[e.g.,][]{lewin93}, assuming a helium atmosphere and canonical 
NS parameters (1.4 solar mass and radius of 10 km), leads to a source
distance of 6.4 kpc. However, throughout
the paper we use the 7.3 kpc observably derived source distance. This value is
in the range 5--8 kpc inferred with the 2000 burst assuming an
hydrogen or helium burst, respectively \citep{intZ02}. 

The best fit to the broadband 3--100 keV persistent emission spectrum of 
\SLX\ required a simple power-law model with a $\Gamma\sim 2.1$.  
This spectral characteristic is similar to those observed in the low/hard 
state of LMXB \citep[see, e.g.,][]{barret00, MF06}. 
Assuming a distance of 7.3 kpc in the direction of 
the Galactic center for \SLX, the estimated persistent unabsorbed flux between
0.1--100 keV, $F_{\rm pers}\approx~3\times 10^{-10}$ erg cm$^{-2}$ s$^{-1}$, 
translates to a bolometric luminosity $L_{\rm pers}\approx~1.9\times
10^{36}$ erg s$^{-1}$ or $\approx0.5\% L_{\rm Edd}$. 
This value is consistent with the value reported by
\citet{intZ02} and  the fact that during the \I\ observation
the source flux slightly increased (see Fig. \ref{fig:lc-rxte}). 
This makes \SLX\ another member of the class of bursters with  
low persistent emission \citep[see, e.g.,][and references
  therein]{coc01,cor04}. 
The mass accretion rate per unit area of the persistent emission,
given by $L_{\rm pers}\;\eta^{-1}$ c$^{-2}/A_{\rm acc}$ (where $A_{\rm
  acc}= 4\pi R_{\rm NS}^{2}$ and $\eta \simeq~0.2$ is the
accretion efficiency for a 1.4 M$_{\odot}$ and 10 km radius NS),
is $\dot m=840$ g cm$^{-2}$ s$^{-1}$. 
Since the {\it RXTE} light curve does not indicate strong differences in the
persistent flux of the source at the time of the four bursts, we are
not able to comment on the exact accretion state at any time. 
We assume, therefore, that the bursts occurred at about the same accretion rate.

\begin{table*}[ht]
\caption{Properties of the most powerful intermediate long bursts.}
\renewcommand{\tabcolsep}{0.2pc} 
\begin{tabular}{lcccccccccc}
\hline\hline
\noalign{\smallskip} 
source  & \multicolumn{4}{c}{SLX1737-282} & SLX 1735-269 & 2S 0918-549 &
IGR J17254-3257 & GX 3+1 & GX 17+2 & 4U 1708-23 \\
\hline
\noalign{\smallskip}
instrument  & WFC & \multicolumn{3}{c}{JEM-X} & JEM-X & WFC & JEM-X  & JEM-X & PCA & SAS-3\\
precursor burst &  no &no&yes&no & yes & no & no & no & no & yes \\
duration (min) & 15 & 25& 30 &20 & 33 & 40 & 15 & 30 & 15--30 & 25 \\ 
$\tau_{\rm rise}$ (sec) & 1 &2&2 &2 & 100 & 1  & 20 & 1.3 & 0.4--1.3 & 20 \\
$\tau_{\rm exp}$ (min)  & 10 &5.0 &5.4 & 4.5 & 10 & 3.9 & 3.7 & 10.8 & 3.2--8.3 & 5.5\\
$kT_{\rm max}$ (keV)  & 3.0 & 3.0 & 2.8&2.3 & 2.9 & 3.0 & 1.6 & 2.3 & 1.8--2.4 & 2.5 \\
$L_{\rm peak}^{a}$ (10$^{38}$\,erg\,s$^{-1}$)  & 3.8&2.5&3.8&3.6 & 5.1 &
3.5 & 0.9 & 0.8 & 1.6--2.0 & 3.0\\ 
$E_{\rm b}$ (10$^{40}$\,erg) & 19&7&12&10 & 20 & 9 & 2.0 & 2.1 &
5.1--7.9 & 9.7 \\
$\tau$$\equiv$E$_{\rm b}$/L$_{\rm peak}$ (min) & 8.4&4.7&5.4&4.7 & 6.5 &
4.3 & 3.6 & 4.4 & 5.3--6.6 & 5.4\\ 
$L_{\rm pers}^{b}$ (\%L$_{\rm Edd}$) & 0.4&0.5&0.5&0.5 & 1.0 & 0.6 &
0.2 & 6.0 & 75--80 & 0.3 \\ 
distance (kpc) & 8 &7.3& 7.3 & 7.3&  8.5 &  5.4 & 8 & 5 & 10 & 6\\
references & [1] &[2]&[2] & [2] &[3] & [4] & [5] & [6] & [7,8,9] &[10] \\ 
\hline
\noalign{\smallskip}
\multicolumn{11}{l}{\footnotesize $^a$\,Unabsorbed bolometric peak
  (black-body) luminosity.} \\ 
\multicolumn{11}{l}{\footnotesize $^b$\,We used the bolometric
  unabsorbed flux from spectral fits; the observed maximum flux during
  radius-expansion bursts} \\ 
\multicolumn{11}{l}{\footnotesize 1. \citet{intZ02}, 2. this work
  3. \citet{molkov05}, \citep[see also][]{suzuki05,sguera07b},}\\
\multicolumn{11}{l}{\footnotesize  4. \citet{intZ05},
  5. \citet{chenevez07}, 6. \citet{chenevez06},
  7. \citet{tawara84}, 8. \citet{kul02},}\\
\multicolumn{11}{l}{\footnotesize 9. \citet{galloway06},
  10. \citet{Hoff}}\\
\end{tabular}
\label{table:lb}
\end{table*}

\subsection{Recurrence time of the bursts}
\label{sec:burst_recurrence_time}

Up to now only four intermediate long bursts, all similar and longer than 15
minutes, have been observed from the low persistent LMXB \SLX\ (see
Table \ref{table:lb}).
The frequency of the bursts observed with \I,
i.e., the total exposure time of $\approx~22.3$ Ms for \SLX\
divided by three, (the number of observed bursts), gives 86 days.
Bursts 1 and 2 have the lowest and highest total energy release:
  $E_{b,1}\simeq 0.7\times 10^{41}$ erg and $E_{b,2}\simeq 1.23\times
  10^{41}$ erg, respectively (at 7.3 kpc distance). This corresponds to an
ignition column $y = E_{b}(1+z)/4\pi R_{\rm NS}^{2} Q_{\rm nuc}$,   
ranging between $y_{1,2}\approx~1.7-2.9\times10^{9}$ g cm$^{-2}$ for
burning hydrogen with abundance X=0.7,
and $y_{1,2}\approx~4.6-8.0\times10^{9}$ g cm$^{-2}$ for X=0 (pure helium);
here $Q_{\rm nuc} = 1.6 + 4X$ MeV nucleon$^{-1}$ 
is the nuclear energy release for a given average hydrogen fraction  
at ignition X, and $z=0.31$ is the 
appropriate gravitational redshift at the surface of a 1.4 M$_{\odot}$ and
$R = 10$ km NS.
Using the relation $\Delta t_{\rm rec}= y_{1,2}(1+z)/\dot m$ a burst
recurrence time of $\approx~31-52$ days is expected for X=0.7, 
and $\Delta t_{\rm rec}\approx~83-144$ days for pure helium burning.
These values are also within the recurrence time range proposed by
\citet{intZ07}. The apparent recurrence time derived from our
\I\ observations is, therefore, more consistent with a pure helium
burning regime. At this quite low accretion rate,
hydrogen ignition would otherwise be likely \citep[see,
  e.g.,][]{boirin07}. However, observation of photospheric radius
expansion is evidence for He-burning.  
We can also
calculate the ratio of the total energy emitted in the persistent flux
to that emitted in  the burst  $\alpha= (F_{\rm pers}/f_{b})\Delta
t_{\rm rec} = (\gamma / \tau) \Delta t_{\rm rec} = 115-203$
for $\Delta t_{\rm rec}= 86$ days and F$_{\rm pers}=3\times10^{-10}$
erg cm$^{-2}$ s$^{-1}$
and $f_{b}$ between $1.1-1.94\times10^{-5}$ erg cm$^{-2}$.  
Assuming again that all the accreted
fuel is burned during the burst, the calculated $\alpha$-value from the
measurable quantities is consistent with a helium burst for $\alpha= 44 M_{\rm
 1.4M\odot}R_{\rm 10 km}^{-1} (Q_{\rm nuc}/{\rm 4.4 MeV/nucleon})^{-1}=121$.
Also in this case, a possible explanation for such an intermediate
long bursting only system could be pure He burning resulting from the long
accumulation of He at low accretion rate, possibly from a pure He donor
companion \citep{cum06}. 

\begin{figure}[ht!]
\centerline{\epsfig{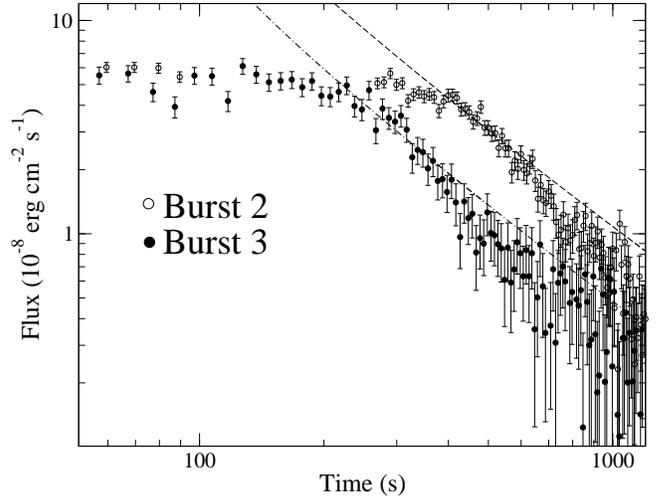}} 
\caption 
{Comparison of the observed decay  of the bolometric black
body flux (open circle burst 2 and circle burst 3) with a theoretical model
\citep{cm04} for the cooling rate of a column of depth 
$7\times10^{9}$~g~cm$^{-2}$ and a nuclear energy release of
$1.6\times10^{18}$~erg~g$^{-1}$, which is expected for helium burning
to iron. The model is shown in dashed and dot-dashed curve for
bursts 2 and 3, respectively. 
}  
\label{fig:fig9} 
\end{figure}

The burst light curves are
consistent with an ignition column depth of
$7\times 10^9\ {\rm g\ cm^{-2}}$. Figure \ref{fig:fig9} compares the burst
light curves with the cooling model of \citet{cm04} in
which an energy of $1.6\times 10^{18}\ {\rm erg\ g^{-1}}$ is deposited
throughout a column of $7\times 10^9\ {\rm g\ cm^{-2}}$, followed by the
cooling of the layer. The models are not valid for
the Eddington limited part of the bursts, since they do not include
the effects of radius expansion, and, therefore, allow super-Eddington
luminosities. However, at late times, the models predict that the flux
should decay as a power-law in time.
Indeed, Fig. \ref{fig:fig9} shows that for 
times $\gtrsim 300$ and $\gtrsim 400$ seconds, following the start of
burst 2 and 3, respectively, the observed decay is a power-law. The
power-law fit is statistically preferred over an exponential fit by
factor 1.1 and 1.5 in the $\chi^{2}_{\rm red}$ for burst 2 and 3, respectively.
The fitted power-law indices for bursts
2 and 3 are $-2.16\pm0.18$. This is a steeper decay than in the models, which
have $F_{\rm Model}\propto t^{-1.55}$ at late times. Further investigation is
needed, but this may reflect a different dependence of the
conductivity on depth, or a different initial temperature profile than
assumed in the models. In Fig. \ref{fig:fig9}, we have adjusted the
normalization 
of the light curves in each case to match the observed decay. The
difference in normalization factors is 2.2. Another way to explain the
different decays is that the column depths of the two bursts are
different, since the time at which the light curves turn over into the
power-law decay depends directly on the thickness of the layer
\citep{cm04}. In that case, burst 3 may have an ignition
depth of about a factor of two smaller than burst 2.

\subsection{Intermediate long bursts: nuclear burning scenarios}
\label{sec:He}

For a given source, the duration of the bursts depends upon the
nature of the donor star and the recurrence time between the different
bursts, i.e. the composition of the accreted fuel and its amount. 
For the high persistent sources the intermediate long bursts may be
produced either by the unstable burning of a large pile of mixed hydrogen 
and helium, where the beta decay of the CNO cycle and rp-process is
responsible for the long 
duration of the bursts, or from the ignition of a thick pure helium layer
 accumulated by the steady burning of hydrogen into helium. In
the case of intermediate long bursts from low persistent bursters, they may be 
produced directly by the slow accretion from a pure helium
donor star \citep{cum06} or if a series of weak hydrogen flashes
generates a massive layer of helium that eventually ignites in an
energetic pure helium flash \citep{Peng,CoopN}.   

The low-persistent source \SLX\ is most likely an ultra-compact X-ray binary 
system with a helium donar star  \citep[see][and Sec. \ref{sec:UCXB}]{intZ07}. 
In this picture, the intermediate long bursts from \SLX\ could be associated
with the pure helium flash regime accreted from a helium donor star
\citep{cum06}. 
Also, the observed burst properties from \SLX\ are consistent
with pure helium ignition at the column depth of 
$\approx5-8 \times 10^{9}$ g cm$^{-2}$,
leading to a burst energy release of $10^{41}$ erg and
a recurrence time of $\approx 130$ days at 0.5\% Eddington luminosity.
To get ignition of helium at a depth
  $\approx5-8\times 10^{9}$ g cm$^{-2}$ requires that the  
heat flux from deeper in the star (which heats the accumulating  
helium layer) is equivalent to approximately 1 MeV/nucleon at 1\%  
Eddington accretion rate \citep{cum06}. Figures 18 and 19 in \citet{cum06} 
show that this is exactly what we expect for models with  
slow modified URCA-like cooling in the NS core. 
Including our recurrence time of \SLX\ on those figures would suggest  
a hot rather than a cold neutron star core.

The composition of the accreted material in intermediate long burst
sources also showing short bursts may not necessarily be pure helium
\citep{intZ07,chenevez07}. 
The presence of some hydrogen may indeed explain the differences 
that distinguish the short from intermediate long bursts.
If a burster source exhibits both
 intermediate long bursts and short bursts, then a fine tuning of the
 accretion rate between the two regimes could be at work near the
 value where the accumulating hydrogen 
transitions from unstable burning at low accretion rates to stable
burning (via the Hot CNO cycle) at higher accretion rates
\citep[e.g.,][]{sb06}.  When stable, the hydrogen burning steadily
accumulates a thick helium shell that eventually ignites. Pure helium
bursts of such thick columns lead to intermediate long burst durations
\citep[e.g.,][]{cum06,Peng}.   
Moreover, \citet{CB00} showed that for pure helium ignition, the
 ignition column is very sensitive to the accretion rate. In
 particular, the transition to unstable hydrogen burning can be quite
 sudden, leading to short mixed H/He bursts. As shown by \citet{CoopN}
 both energetic pure helium 
flashes and weak hydrogen flashes may occur near the transition. These
weak hydrogen bursts (undetectable because their peak luminosity is
lower than the persistent luminosity) contribute to the building 
of the deep layer of nearly pure helium. Such weak bursts may also
they trigger the ignition of the helium if its mass is sufficiently
large. So, the intermediate long bursts result from the ignition of a
large helium pile beneath a steady hydrogen burning shell.

\subsection{Comparison to other burster sources}
 \label{sec:comparison}

\begin{figure}[htb!]
\centerline{\epsfig{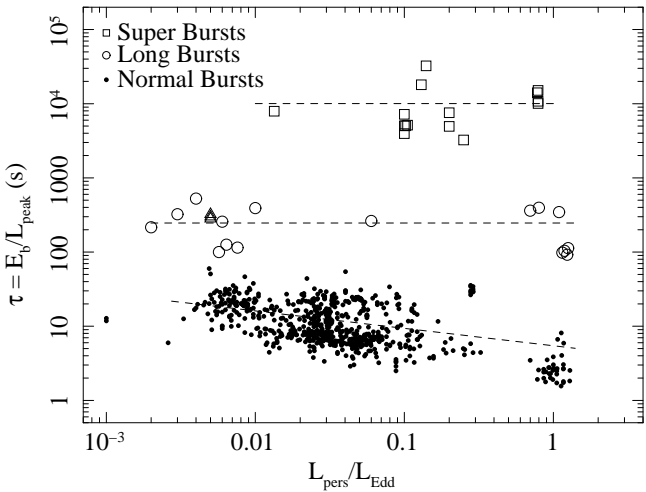}} 
\caption 
{Bursts effective durations vs. persistent luminosity for normal
  bursts (points)
observed with {\it RXTE}  \citep[see][]{galloway06}, intermediate long
bursts  \citep[open circle, and triangle for this work;  see Table
  \ref{table:lb} and for the 
  $\tau\approx$ 100 s bursts see][]{galloway06}, and Superbursts 
 \citep[open square, see e.g.,][see also
 Sect. \ref{sec:comparison}]{ek04,intZ04}.}  
\label{fig:fig6} 
\end{figure} 

The physics of the time dependent type I X-ray bursts is
associated with the thermal unstable thermonuclear reactions
\citep[e.g.,][]{sb06}. Depending on different local accretion
rates, the amount of fuel, and the type of nuclear burning, we can distinguish
three main burst branches, normal bursts, intermediate long bursts, and
superbursts (see Fig. \ref{fig:fig6}). We defined the different bursts
type in  Fig. \ref{fig:fig6} as follows: normal bursts are distributed
along a power-law fit with index $\Gamma=-0.24$; the intermediate long
bursts with a duration of $\tau=250$ s and the superbursts with
$\tau=2.8$ hr. Most of the short bursts are observed at accretion rates between
0.005--0.2$\dot M_{\rm Edd}$  or around the Eddington limit. For the
low accretion rates ($\dot M < 0.01\dot M_{\rm Edd}$), these bursts are
mixed H/He burning triggered by thermally unstable H ignition. For
intermediate accretion rates ($0.01\dot M_{\rm Edd} \le \dot M \le
0.02-0.07\dot M_{\rm Edd}$) they are pure He shell ignition after
steady H burning, and for high accretion rates  ($0.02-0.07\dot M_{\rm
  Edd} \le \dot M \approx 1\dot M_{\rm Edd}$) they are H/He burning
triggered by thermally unstable He ignition \citep[see, e.g.,][and
  references therein]{sb06}. Note that in Fig. \ref{fig:fig6} we also
have the relation $L_{\rm pers}/L_{\rm Edd} \simeq \dot M_{\rm
  acc}/\dot M_{\rm Edd}$.  

All sources showing intermediate long bursts or superbursts also exhibit
normal bursts, except \SLX. Sofare, the latter has only showed
intermediate long bursts. In Table \ref{table:lb}, we report the
properties of the most powerful and recently studied intermediate long
bursts. In Fig. \ref{fig:fig6}, we also added intermediate long
bursts with $\tau\approx100$ s from the following sources:
GRS 1747-312, EXO 0748-676, and GX 17+2 \citep{intZ03,galloway06}.
The burst properties of \SLX\ are similar to the other 
intermediate long bursts.
The intermediate long burst from SLX 1735-269 \citep{molkov05} is the only
one showing a remarkably long rise time. This was due to an extended
photospheric radius expansion phase with a well separated
precursor.  \citet{molkov05} interpreted the long decay as 
most probably due to the mixed burning of H/He. However,
\citet{intZ07} suggested this source is an ultra-compact X-ray
binary system and, therefore, a pure helium burst cannot be ruled out at
this relatively low accretion rate. 
Note that only the two high accretion rate
sources, GX 3+1 and GX 17+2, show all three kinds of bursts,
 the latter being  the only source that shows intermediate long
  bursts at the Eddington mass accretion rate \citep[most likely
    resulting from mixed burning of H/He,]
  []{chenevez06,galloway06,kul02,intZ04,kuulers02}.   

The superbursts
are observed between 0.1--0.3$\dot M_{\rm Edd}$ and the intermediate
long bursts are observed between 0.002--0.01$\dot M_{\rm Edd}$, except
again for GX 17+2 at $\sim 1\dot M_{\rm Edd}$ and GX 3+1 for the peculiar
two-phase intermediate long
burst at $\sim 0.06\dot M_{\rm Edd}$ (see Table \ref{table:lb}). 
For Fig. \ref{fig:fig6}, we derived the persistent bolometric
luminosity and burst duration for the superbursts 4U 0614+09
\citep{ek05} and 4U 1608-52 \citep{remillard05} and found $L_{\rm
  pers} \approx 0.013$ and $0.14L_{\rm Edd}$ and $\tau =
0.15\times10^{42}/0.2\times10^{38} \approx 2.01$ hr and  $\tau
=2\times10^{42}/0.6\times10^{38} < 9.2$ hr, respectively \citep[see
 also for 4U 1608-52][]{keek07}. 
For the first time a superburst, from 4U 0614+09, has been
observed at  $\sim0.01 \dot M_{\rm Edd}$ mass accretion rate, which
diverges from the current prediction that superbursts with carbon
ignition on the hot NS crust require an accretion rate $ > 0.1\dot
M_{\rm Edd}$ \citep[see, e.g.,][]{sb06}. A consideration of these
observations will be presented in \citet{Kuulkers08}.
However, one puzzling issue is to understand why some sources undergo
the three types of bursts and others only one or two.

\subsection{\SLX\ an ultra-compact X-ray binary system}
 \label{sec:UCXB}

\begin{figure}[htb!]
\centerline{\epsfig{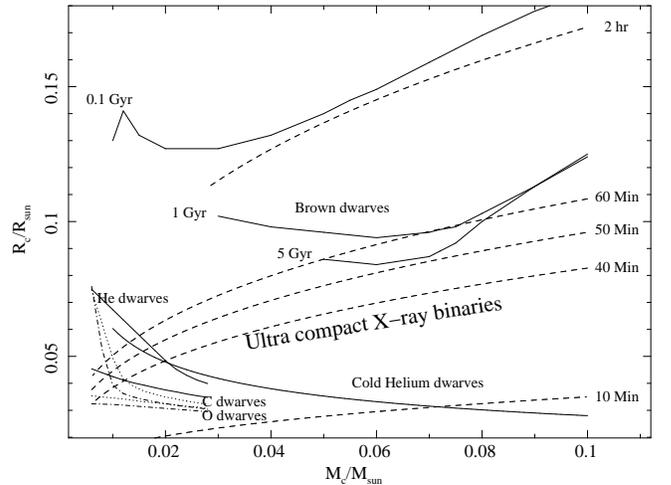}} 
\caption 
{Companion radius $R_{\rm c}$ vs. mass $M_{\rm c}$ plane, showing the
  Roche lobe constrains for the ultra-compact X-ray binaries, for
  $M_{\rm NS}=1.4 M_{\odot}$. The equations of
  state are shown for brown dwarves (solid line)  and cold pure
  helium dwarves. The brown dwarf models are shown for different ages.
The figure also shows the low-mass
  regime for degenerate dwarf models incorporating different compositions
  (dot-dash O, dotted C, line He) and low ($10^{4}$ K) or high
  ($3\times 10^{4}$ K) central temperatures (lower and upper curves).
}  
\label{fig:fig7} 
\end{figure} 

\begin{figure}[htb!]
\centerline{\epsfig{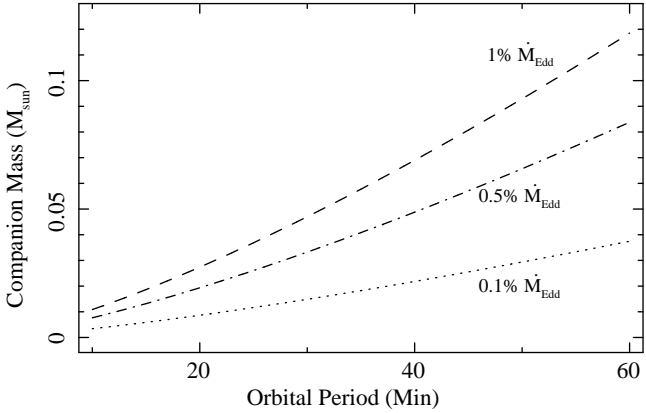}} 
\caption 
{Companion mass $M_{\rm c}$ vs. orbital period for different mass
  accretion rates, assuming that the mass transfer for UCXBs is driven
  by gravitational radiation, i.e., $<\dot M> = \dot M_{\rm GR}$ The
  companion star mass is calculated for 1\%, 0.5\%, and 0.1\% $\dot M_{\rm
  Edd}$ mass accretion rates for orbital periods between 10--60 minutes.
}  
\label{fig:fig8} 
\end{figure}

The source \SLX\ has recently been proposed to be an 
ultra-compact X-ray binary (UCXB) candidate, suggesting a
pure He white dwarf donor star. An UCXB with a hydrogen
poor donor star can sustain persistently low enough accretion rates, 
while a mixed hydrogen/helium accretor with low enough
accretion rates may not exist since they then turn to be transient
\citep{intZ07}. 

To accrete matter persistently, the assumption of a Roche lobe-filling
companion \citep{p71} implies the mass-radius relation, $R_{\rm lobe}=R_{\rm c} =
1.524\times10^{-2}(M_{\rm c}/1M_{\odot})^{1/3}(P_{\rm orb}/1 {\rm
  min})^{2/3}\, R_{\odot}$, shown in Fig. \ref{fig:fig7} for different
orbital periods. We divided the 
orbital periods into two distinct ranges - either around 10--60 -minute
for UCXB or $>2$ hours. In the orbital period regions of 10--60 minutes,
only very low-mass degenerate O, C, or He dwarves can be the donor
star. Recent models of low-mass degenerate dwarves have been produced
incorporating the effect of different compositions and temperatures
\citep{db03}. The corresponding $R_{\rm c}$ versus $M_{\rm c}$
equations of state are also shown in Fig. \ref{fig:fig7}. 
We are particularly interested in this region to study the properties
of \SLX\ most likely to be a pure He accretor.  An additional constraint of
the donor star is given by the minimum accretion rate driven by
gravitational radiation in a close binary \citep[see][for a review]{vh95}. Following
\citet{bc01} we can constrain the source companion mass as a function
of the mass loss rate due to gravitational radiation, i.e., we set
$0.001-0.01 \dot M_{\rm Edd} = 1.3163\times 10^{-7} (M_{\rm c}/0.01
M_{\odot})^{2}(M_{\rm NS}/1.4 M_{\odot})^{2/3}(P_{\rm orb}/1 {\rm
  min})^{-8/3}$ M$_{\odot}$/yr for conservative mass transfer. Note that the source distance is not a
free parameter since it is determined from the Eddington luminosity
reached during the burst photospheric radius expansion. Figure 8 shows
that for mass accretion rates lower than $0.01\dot M_{\rm Edd}$, a pure helium
donor star can be the companion star for a He only bursting source
\citep[see also][for evolutionary considerations]{intZ05}.
The knowledge of the orbital period and mass function of
the system could more strongly constrain the companion star and
evolution of the system, see, e.g., for the UCXB accreting X-ray millisecond
pulsars like XTE~J1751-305 \citep{markwardt02} or XTE~J1807-294
\citep{MF05}. 
Given the mass function, f(M)= (M$_{\rm c}\, sin\,i)^{3}$/(M$_{\rm c}$ + M$_{\rm
NS})^{2}$, an important free parameter is the inclination
angle of the system. Assuming an inclination angle of the binary
system between the mean value  of 60$^{\circ}$  to a maximum value of
$\sim85^{\circ}$  the 
donor star has to be between 0.005 and 0.03 M$_{\odot}$. 
This requires a low persistent mass accretion rate
of $<0.01 \dot M_{\rm Edd}$ as shown in Fig. \ref{fig:fig8}. 

In the $>2$ hr orbital period regions the donor star can be a hydrogen  
main-sequence star or a brown dwarf. The superburst sources belong to
this region except 4U 1820-303, which has an orbital period of
$\sim 11.5$ minutes. Most likely at variance with the UCXB intermediate
long bursters,  4U 1820-303 has a very low inclination angle of between
30--35$^{\circ}$ and a higher persistent emission, therefore, a
more massive donor star, see Figs. \ref{fig:fig7} and \ref{fig:fig8}.   
Also, transient accreting millisecond pulsars in this
orbital period range, like SAX J1808.4-3658
or HETE J1900.1-2455 show  short bursts \citep{gc06,fal07}. For
these sources, the time between bursts was long enough for hot CNO
burning to significantly deplete the accreted hydrogen, so that
ignition occurred in a pure helium layer underlying a stable hydrogen
burning shell.

\section{Conclusions}
\label{conclusions}
 
From an observational stand point, \SLX\ is the only 
source that  exclusively exhibits intermediate long bursts lasting about
20-30 minutes. These intermediate long bursts are most likely powered
by the burning of a thick He layer. This is
a consequence of the low accretion rate coupled most likely with the
ultra-compact X-ray binary nature,  which is also consistent with
the lack of observed soft short bursts for  \SLX. The observed properties of the
helium bursts support these conclusions, also requiring $\simeq90$ days
recurrence time.

\acknowledgements 
MF acknowledges the French Space Agency (CNES) for financial support.
JC acknowledges financial support from ESA-PRODEX, Nr. 90057.

\end{document}